\documentclass[12pt]{article}
\usepackage{amsmath,amsfonts,epsfig}
\usepackage{braket}
\usepackage[usenames]{color} %%%%%%%%%%%
\usepackage{wrapfig}
\usepackage{amsmath,amsfonts,amssymb,bm}
\usepackage{amsfonts,bm}
\usepackage{tikz}
\usepackage{mathtools}
\newcommand{\p}{\partial}
\newcommand{\pslash}{p\kern-1ex /}
\newcommand{\qslash}{q\kern-1ex /}
\newcommand{\lslash}{l\kern-1ex /}
\newcommand{\sslash}{s\kern-1ex /}
\newcommand{\kaslash}{k_a\kern-2ex /}
\newcommand{\kbslash}{k_b\kern-2ex /}
\newcommand{\Dslash}{{\cal D}\kern-1.5ex /}

\newcommand{\bc}{\overline{c}}

\newcommand{\beqa}{\begin{eqnarray}}
\newcommand{\eeqa}{\end{eqnarray}}

\newcommand{\bpm}{\begin{pmatrix}}
\newcommand{\epm}{\end{pmatrix}}
\newcommand{\bbm}{\begin{bmatrix}}
\newcommand{\ebm}{\end{bmatrix}}

\def\p{\partial}

\makeatletter

\@addtoreset{equation}{section}
\makeatother
\usepackage{subfig}
\begin{document}

%\voffset -0.7 true cm
%\hoffset 1.1 true cm
%\topmargin 0.0in
%\evensidemargin 0.0in
%\oddsidemargin 0.0in
%\textheight 8.6in
%\textwidth 7.1in
%\parskip 10 pt

\voffset -0.7 true cm
\hoffset 1.5 true cm
\topmargin 0.0in
\evensidemargin 0.0in
\oddsidemargin 0.0in
\textheight 8.6in
\textwidth 5.4in
\parskip 9 pt
 
\def\Tr{\hbox{Tr}}
\newcommand{\be}{\begin{equation}}
\newcommand{\ee}{\end{equation}}
\newcommand{\bea}{\begin{eqnarray}}
\newcommand{\eea}{\end{eqnarray}}
\newcommand{\beas}{\begin{eqnarray*}}
\newcommand{\eeas}{\end{eqnarray*}}
\newcommand{\nn}{\nonumber}
\font\cmsss=cmss8
\def\C{{\hbox{\cmsss C}}}
\font\cmss=cmss10
\def\bigC{{\hbox{\cmss C}}}
\def\scriptlap{{\kern1pt\vbox{\hrule height 0.8pt\hbox{\vrule width 0.8pt
  \hskip2pt\vbox{\vskip 4pt}\hskip 2pt\vrule width 0.4pt}\hrule height 0.4pt}
  \kern1pt}}
\def\ba{{\bar{a}}}
\def\bb{{\bar{b}}}
\def\bc{{\bar{c}}}
\def\bphi{{\Phi}}
\def\Bigggl{\mathopen\Biggg}
\def\Bigggr{\mathclose\Biggg}
\def\Biggg#1{{\hbox{$\left#1\vbox to 25pt{}\right.\n@space$}}}
\def\n@space{\nulldelimiterspace=0pt \m@th}
\def\m@th{\mathsurround = 0pt}

\begin{titlepage}
\begin{flushright}
{\small OU-HET-1145}
 \\
\end{flushright}

\begin{center}

\vspace{5mm}
{\Large \bf {Flows of Extremal Attractor Black Holes}} \\[3pt]

\vspace{6mm}

\renewcommand\thefootnote{\mbox{$\fnsymbol{footnote}$}}
Norihiro Iizuka${}^{1}$, 
Akihiro Ishibashi${}^{2}$ and 
Kengo Maeda${}^{3}$

\vspace{3mm}

${}^{1}${\small \sl Department of Physics, Osaka University} \\ 
{\small \sl Toyonaka, Osaka 560-0043, JAPAN}

${}^{2}${\small \sl Department of Physics and Research Institute for Science and Technology,} \\   
{\small \sl Kindai University, Higashi-Osaka 577-8502, JAPAN} 
%\\ 

${}^{3}${\small \sl Faculty of Engineering, Shibaura Institute of Technology,} \\   
{\small \sl Saitama 330-8570, JAPAN} 
%\\ 

\vspace{4mm}

{\small \tt 
{iizuka at phys.sci.osaka-u.ac.jp}, {akihiro at phys.kindai.ac.jp},   \\
{maeda302 at sic.shibaura-it.ac.jp}
}

\end{center}

\vspace{3mm}

\noindent

\abstract{
We study flows of non-supersymmetric attractor black holes in the context of gauge/gravity correspondence. As our bulk theory, we consider the Einstein-Maxwell-Dilaton system with a single dilaton field coupled to two Maxwell fields and make a relevant deformation by adding a bare potential to the dilaton field. We find two types of extremal black hole solutions with attractor mechanism: The one smooth at the horizon and the other non-smooth. We show from both bulk and boundary theory perspective that the former is thermodynamically unstable, while the latter is stable. 
}

\end{titlepage}

\setcounter{footnote}{0}
\renewcommand\thefootnote{\mbox{\arabic{footnote}}}

\tableofcontents
\newpage
%%%%%%%%%%%%%%%%%%%%%%%%%%%%%%%%%%%%%%%%%%%%%%%%%%%%%%%%%%%

%%%%%%%%%%%%%%%%%%%%%%%%
\section{Introduction} 
%%%%%%%%%%%%%%%%%%%%%%%%

Extremal black holes show a very interesting phenomena, called {\it attractor mechanism}, where several moduli fields are drawn to fixed values at the black hole horizon and those values are determined only by the charges of the black holes. 
Historically attractor mechanism was first found for BPS black holes in ${\cal{N}} =2 $ supergravity in 4-dimension \cite{Ferrara:1995ih}, and studied extensively in '90 by  \cite{Cvetic:1995bj,Strominger:1996kf,Ferrara:1996dd,Ferrara:1996um,Cvetic:1996zq,Ferrara:1997tw,Gibbons:1996af,Denef:2000nb,Denef:2001xn}.  Later in '00, it is revisited and pointed out in \cite{attractor} that this phenomena appears not only in BPS black holes, but also in more generic settings as long as black holes are extremal ({\it i.e.,} zero temperature limit), and their ``effective potential'' satisfies certain criteria. It is also pointed out by Sen in \cite{Sen:2005wa} (see the lecture note \cite{Sen:2007qy} as well) that on the situation where attractor mechanism appears, the fixed moduli values at the horizon are determined as extremal values for the entropy function.  Attractor for extremal rotating black holes was also studied 
\cite{Astefanesei:2006dd}.

Besides attractor mechanism, extremal black holes are quite interesting by their own. One of their peculiar nature is their nonzero horizon areas, which correspond to non-zero entropies and therefore signaling their large degeneracy in zero temperature limit. Through the gauge/gravity duality, an extremal black hole corresponds to highly degenerate ground states of the dual field theory which has finite charge density. Since ground states play a crucial role in physics, it is important to classify these highly degenerate ground states from the bulk dual, which naturally leads us to classifying various extremal black holes. One such classification is Bianchi classification of extremal black holes \cite{Iizuka:2012iv,Iizuka:2012pn} which lead to generic Bianchi type of extremal black holes possibly dual to some condensed matters. See for examples, \cite{Donos:2012gg,Donos:2012wi,Iizuka:2013wya}. 

Another interesting aspect of extremal black holes is that one can freely tune the moduli value both at the boundary and also at the horizon. This might make it possible to study the nature of extremal black holes in terms of the boundary theory since the boundary value of the bulk moduli corresponds to the coupling of the boundary theory through holography \cite{Dabholkar:2006tb}. Another interesting point of extremal black hole is its apparent stability, and the stability of the extremal black holes cause phenomenological puzzle which lead to the weak gravity conjecture \cite{Arkani-Hamed:2006emk}. 
In these ways, extremal black holes are key objects in understanding various aspects of gravity and also gauge/gravity correspondence.

Going back to the holographic interpretation of extremal black holes as degenerate ground states, it is interesting to see how these degenerate vacua change by introducing relevant operators through an RG flow. 
In this paper, we investigate this question from the bulk dual. The question we ask in this paper is the following; 
{\it Given the non-supersymmetric extremal black holes which show attractor mechanism in asymptotic AdS spacetime, once we introduce a relevant deformation to this bulk theory through a bare potential, how does bulk extremal black hole flow?}

The organization of this paper is as follows. 
In section \ref{sec:2}, we review non-supersymmetric attractors for extremal black holes in asymptotic AdS$_4$ background.  
Then in section \ref{sec:3}, we analyze the effects of a relevant operator in the boundary theory from the bulk, which is induced by adding a bare potential for the dilaton (moduli) in the bulk. The bare potential induces flows from a non-supersymmetric but extremal attractor black hole to another extremal black hole. However the effects of bare potential induces critical difference for the extremal black hole near the horizon, which is analyzed in detail in section \ref{sec:3}.  
In section \ref{sec:4}, we discuss stability of our new extremal black holes. Section \ref{sec:5} is for summary and discussions.  
 
{Before we close this introduction, we comment on several literatures. 
Recently the RG flow of the boundary theory was studied vigorously from the dual bulk by adding relevant operators in \cite{FHKS2020,HHKS2020, HHKS2021}. Especially in \cite{HHKS2020, HHKS2021}, the singularity of would-be Cauchy horizon was studied in detail and curious scaling region was found between outer horizon and inner horizon. 
On the other hand, for the extremal black hole case which we study in this paper, inner and outer horizon always coincides and there is no region in between. } 

%%%%%%%%%%%%%%%%%%%%%%%%
\section{Attractor mechanism in AdS$_4$}
\label{sec:2} 
%%%%%%%%%%%%%%%%%%%%%%%%

\subsection{The setup}

The model we consider is the Einstein-Maxwell-dilaton theory where we have two species of $U(1)$ fields, $F_{\mu\nu}$ and $H_{\mu\nu}$ with a cosmological constant $\Lambda$; 
\begin{align}
\label{actionII} 
& S=\int d^4x\sqrt{-g}\left[R-2(\nabla\phi)^2-e^{2 a \phi}F^2-e^{-2 a \phi}H^2
-  2 \Lambda 
  \right], \nonumber \\
& F_{\mu\nu}=\p_\mu A_{\nu}-\p_\nu A_{\mu}, \qquad H_{\mu\nu}=\p_\mu B_{\nu}-\p_\nu B_{\mu} . 
\end{align}
This is a typical model which shows non-supersymmetric attractor mechanism \cite{attractor} for the case of vanishing cosmological constant,  $\Lambda = 0$. 

The equations of motion for the dilaton $\phi$, the Maxwell fields $F_{\mu\nu}$, $H_{\mu\nu}$, and the metric $g_{\mu\nu}$ are 
\begin{align}
\label{basic_Eqs1}
 \Box\phi &=\frac{1}{2} a  (e^{2 a \phi}F^2-e^{-2 a \phi}H^2),   \\
 \label{basic_Eqs2}
 \nabla_\nu(e^{2 a \phi} F^{\nu\mu}) &=\frac{1}{\sqrt{-g}}\p_\nu(e^{2 a \phi}\sqrt{-g}F^{\nu\mu})=0,   \\
 \label{basic_Eqs3}
 \nabla_\nu(e^{-2 a \phi} H^{\nu\mu}) &=\frac{1}{\sqrt{-g}}\p_\nu(e^{-2 a \phi}\sqrt{-g}H^{\nu\mu})=0,   \\
 \label{basic_Eqs4}
 R_{\mu\nu} & =2\nabla_\mu\phi\nabla_\nu \phi+2(e^{2 a \phi}F_{\mu\alpha}{F_\nu}^\alpha 
+e^{-2 a \phi}H_{\mu\alpha}{H_\nu}^\alpha)
\nonumber \\
 & \qquad \qquad \qquad +\frac{1}{2}g_{\mu\nu}( 2 \Lambda -e^{2 a \phi}F^2-e^{-2 a \phi}H^2). 
\end{align}

We consider the case where the boundary space is a plane $R^2$. 
In such a case, one can always take the following form
\begin{align}
\label{metric_diagonal}
ds^2=\frac{1}{z^2}\left(-f(z)e^{-\chi(z)}dt^2+\frac{dz^2}{f(z)}+dx^2+dy^2    \right)
\end{align}
for the generic static metric, where $z=0$ is the AdS boundary and $f(z) = 0$ represents the black hole horizon. 

The metric eq.~\eqref{metric_diagonal} can also be written by introducing the following null coordinate
\be 
v=t-\int \frac{e^{\frac{\chi}{2}}}{f}dz . 
\ee
Then eq.~\eqref{metric_diagonal} reduces to 
\begin{align}
\label{advanced_null}
ds^2=\frac{1}{z^2}\left(-f(z)e^{-\chi(z)}dv^2-2e^{-\frac{\chi(z)}{2}}dvdz+dx^2+dy^2 \right) \,.
\end{align}
Eq.~\eqref{advanced_null} is well-behaved behind the horizon. 
% than eq.~\eqref{metric_diagonal}. 

Our static flux ansatz is 
\begin{align} 
\label{gauge_ansatz}
A_\mu=(A_v(z), 0, 0, 0), \qquad B_\mu=(B_v(z), 0, 0, 0). 
\end{align}
From the equation of motion for $F_{\mu\nu}$ eq.~\eqref{basic_Eqs2} and $H_{\mu\nu}$ eq.~\eqref{basic_Eqs3}, 
we obtain 
\begin{align}
\label{sol_Av}
 A_v'=Q_A e^{-2 a\phi}e^{-\frac{\chi}{2}}, \qquad B_v'=Q_B e^{2 a\phi}e^{-\frac{\chi}{2}}
\end{align}
where $'$ represents the derivative w.r.t. $z$ and the constants $Q_A$ and $Q_B$ represent the charges of the black hole due to $A_\mu$ and $B_\mu$, respectively. 
By plugging this into the equation of motion for the dilaton eq.~\eqref{basic_Eqs1}, we obtain  
\be 
\label{scalar_field_ccEq}
f\phi''+\left(f'-\frac{f}{2}\chi'-\frac{2f}{z} \right)\phi'+a z^2(Q_A^2e^{-2 a \phi}-Q_B^2e^{2 a \phi})
=0 . 
\ee
As eq.~(21) of \cite{attractor}, it is convenient to define an effective potential for the dilaton as 
\be
\label{defeffectivepotential}
V_{\rm eff} :=  Q_A^2e^{-2 a \phi}+ Q_B^2e^{2 a \phi} . 
\ee
Then, eq.~\eqref{scalar_field_ccEq} reduces to 
\be 
\label{scalar_field_Eq_effectivep}
f\phi''+\left(f'-\frac{f}{2}\chi'-\frac{2f}{z} \right)\phi' = \frac{1}{2} z^2 \partial_\phi V_{\rm eff}(\phi) , 
\ee
and the Einstein equation, eq.~\eqref{basic_Eqs4}, reduces to 
\begin{align}
\label{chi_Eq} 
& \chi'=2z\phi'^2,  \\
\label{f_Eq} 
& f'=z^3 V_{\rm eff} +\frac{ \Lambda}{z}+\frac{f}{z}(3+z^2\phi'^2) \,.
\end{align}

\subsection{Attractor conditions}

As analyzed in \cite{attractor}, the attractor value $\phi_0$ is determined from the effective potential only as  
\be
\label{attractorpoint}
\partial_\phi V_{\rm eff} |_{\phi = \phi_0}= 0  \,,
\ee
and the condition for attractor mechanism is 
\be
\label{attractorcondition}
M^2_0 := \frac{1}{2} \partial^2_\phi V_{\rm eff}  |_{\phi = \phi_0}> 0  \,.
\ee
For the case where the effective potential is given by eq.~\eqref{defeffectivepotential}, 
the attractor condition eq.~\eqref{attractorcondition} is automatically satisfied for any real value of $a$. 
{For the case $Q_A = Q_B$, % since the effective potential is given by eq.~\eqref{defeffectivepotential}, 
the attractor value takes the simplest form as
$\phi_0 = 0$,   
but in general case $Q_A \neq Q_B$, 
$\phi_0 \neq 0$. 
%For the case where the effective potential is given by eq.~\eqref{defeffectivepotential}, 
%the attractor condition eq.~\eqref{attractorcondition} is automatically satisfied for any real value of $a$.  
}

As is shown in \cite{attractor}, the minimal value of the effective potential sets the scale of the horizon. In our coordinate choice,  eq.~\eqref{f_Eq} sets the horizon. To see this, note that attractor mechanism works only for the extremal black holes \cite{attractor}, and therefore at the horizon we have 
\be
f = f' = 0  \quad \mbox{(at the extremal horizon)}
\ee
and from  eq.~\eqref{f_Eq}, we have 
\be
z_h^3 V_{\rm eff}|_{\phi=\phi_0} +\frac{ \Lambda}{z_h}   = 0 . 
\ee
In the unit where we set the AdS length set to be one, 
this fixes the horizon at $z = z_h$ as 
\be
\label{horizoncondition}
z_h =  \left( \frac{- \Lambda}{V_{\rm eff}} \right)^{1/4} |_{\phi=\phi_0}  \,, \quad \Lambda = -3  < 0 \,. 
\ee

\subsection{Perturbative analysis}
Starting with extremal Reissner-Nordstrom black hole as a zero-th order solution, one can obtain 
perturbative analytic attractor solution as follows; 
\begin{align}
\label{phinexpansion}
\phi = \sum_{n=0}^\infty \epsilon^n \phi_n , \\
\label{chinexpansion}
\chi = \sum_{n=0}^\infty \epsilon^n \chi_n , \\
\label{fnexpansion}
f = \sum_{n=0}^\infty \epsilon^n f_n ,   
\end{align}
where $\epsilon$ is set by the leading perturbation by the dilaton (moduli) field fluctuation. 
To show how the perturbation works, only in this subsection, we set the two Maxwell charges to be the same as 
\be
Q_A = Q_B := Q , 
\ee
for simplicity. Since in this case, the attractor value is simply $\phi_0 = 0$, and we have 
\be
V_{\rm eff}|_{\phi = 0} = 2 Q^2 . 
\ee
In this case, the zero-th order solution is the extremal Reissner-Nordstrom black hole solution with the metric 
functions,  
\begin{align}
&\qquad \qquad \quad \quad \chi_0  = 0 \,, \\
& f_0  
= \left( 1 - \frac{z_h}{z} \right)^2 \left( 1 + 2 \frac{z_h}{z}+ 3 \left( \frac{z_h}{z} \right)^2 \right) \,.
\end{align} 
Here we set 
\be
\Lambda = -3  \,, 
\ee
with the trivial dilaton profile
\be
\phi_0 = 0 , 
\ee
and the extremal horizon is at 
\be
\label{zhvalue}
z_h = \left( \frac{3}{V_{\rm eff}}\right)^{1/4}|_{\phi = 0}  =  \left( \frac{3}{2 Q^2}\right)^{1/4}  \,.
\ee
The first order perturbation is given by the scalar perturbation only. This can be understood as that the dilaton perturbation $\epsilon \phi_1$ contributes to the Einstein equation through their stress tensor, which is the order of $O(\epsilon^2)$. This sets 
\be
\chi_1 = f_1 = 0 \,. 
\ee
Then, by plugging eq.~\eqref{phinexpansion} into eq.~\eqref{scalar_field_Eq_effectivep}, the leading order perturbation for $\phi_1$ 
satisfies the following linear equation 
\be 
\label{firstperturbation}
f_0 \phi_1''+\left(f_0'-\frac{f_0}{2}\chi_0'-\frac{2f_0}{z} \right)\phi_1' -  z^2 M_0^2 \phi_1 = 0,  
\ee
where 
\be
\label{Mvalue}
M_0^2 =  \frac{1}{2} \left( \partial^2_\phi V_{\rm eff}(\phi) \right) |_{\phi = \phi_0 = 0} =4 a^2 Q^2.  
\ee
Near the horizon $z = z_h$, we have 
\be
f_0 \simeq  6 \left( \frac{ z - {z_h}}{z_h} \right)^2 \,, \quad \chi_0 = 0 \,, 
\ee
and then eq.~\eqref{firstperturbation} reduces to 
\be
 6 \left( \frac{ z - {z_h}}{z_h} \right)^2 \phi_1'' + 12 \frac{z - z_h}{z^2_h} \phi_1' - z_h^2 M_0^2 \phi_1 = 0. 
\ee
The solution behaves near the horizon as 
\begin{align}
& \qquad \qquad \qquad \phi_1 \approx \left( \frac{z - z_h}{z_h}\right)^{\gamma} \,, \\
\gamma &= \frac{-1 + \sqrt{1 + \frac{2}{3} M_0^2 z_h^4}}{2}  = \frac{-1 + \sqrt{1 + 4 a^2 }}{2}  \, > 0 \,, 
\label{gammadefinition}
\end{align}
where we choose the sign in such a way that the dilaton does not blow up at the horizon. $z_h$ and $M_0$ are given by eqs.~\eqref{zhvalue} and \eqref{Mvalue}. Note that $\gamma$ is a {\it continuous} parameter.  
Then from eq.~\eqref{chi_Eq}, 
the backreaction to $\chi_2$ can be obtained near the horizon as 
\be
\chi_2 \approx \frac{2 \gamma^2}{2 \gamma -1}  \left( \frac{z - z_h}{z_h}\right)^{2 \gamma-1}  \,,
\ee
and from eq.~\eqref{f_Eq}, 
\be
f_2' - \frac{3}{z_h} f_2 \simeq f_2' = z_h^3 M_0 \phi_1^2 + f_0 z_h \phi_1'^2 \simeq
\left( z_h^3 M_0 + \frac{6 \gamma^2}{z_h} \right)  \left( \frac{z - z_h}{z_h}\right)^{2 \gamma} \,.
\ee
This allows the near horizon behavior as 
\be
f_2 \simeq  \frac{\left( z_h^4 M_0 +  6 \gamma^2 \right) }{ (2 \gamma + 1)} \left( \frac{z - z_h}{z_h}\right)^{2 \gamma + 1 } \,,
\ee
which justifies $f_2' \gg f_2$. 

Higher order perturbations can be done as well. 
In general, one can check that the following ansatz satisfies the perturbative solution 
\be
\phi_n \propto  \left( \frac{z - z_h}{z_h}\right)^{n \gamma} \,,
\ee
from \eqref{scalar_field_Eq_effectivep}. It follows from eq.~\eqref{chi_Eq},  
\be
\chi_n  \propto  \left( \frac{z - z_h}{z_h}\right)^{n \gamma -1} \,,
\ee
and from eq.~\eqref{f_Eq},  
\be
f_n  \propto  \left( \frac{z - z_h}{z_h}\right)^{n \gamma +1} \,. 
\ee
In these ways, we can obtain all order solutions by perturbation. 

Since the scalar curvature of the metric \eqref{metric_diagonal} is 
\begin{align}
R = z^2 f(z) \chi ''(z)+\frac{3}{2} z^2 \chi '(z) f'(z)-\frac{1}{2} z^2
   f(z) \chi '(z)^2  \nonumber \\
   -3 z f(z) \chi '(z)-z^2 f''(z)+6 z f'(z)-12 f(z), 
\end{align}
$\gamma$ given by eq.~\eqref{gammadefinition} must satisfy 
\be
\gamma \geq  \frac{1}{2} \,
\ee
to avoid the scalar curvature singularity at the horizon $z=z_h$. 

%%%%%%%%%%%%%%%%%%%%%%%%
\section{Flow of the attractor black holes in AdS$_4$} 
\label{sec:3} 
%%%%%%%%%%%%%%%%%%%%%%%%
\subsection{Relevant deformation by the bulk bare potential $V(\phi)$}
%%%%%%%%%%%%%%%%%%%%%%%%

To add a relevant operator, in the bulk we add the following deformation term of the moduli potential into the bulk action eq.~\eqref{actionII} 
\begin{align}
\label{barepotential1}
\delta S &= \int d^4x  \left(  - V(\phi) + 2  \Lambda  \right) \,.
\end{align}
The Maxwell equations are unchanged from eq.~\eqref{basic_Eqs2} and \eqref{basic_Eqs3}, and only the equation of motion for the dilaton and 
the Einstein equation are modified as 
\begin{align}
\label{basic_Eqs_dilaton}
 \Box\phi & =\frac{1}{2}a (e^{2a \phi}F^2-e^{-2a \phi}H^2)
+\frac{1}{4}\frac{\p V}{\p \phi},   \\
 R_{\mu\nu} & =2\nabla_\mu\phi\nabla_\nu \phi+2(e^{2a \phi}F_{\mu\alpha}{F_\nu}^\alpha
+e^{-2a \phi}H_{\mu\alpha}{H_\nu}^\alpha) \nonumber \\
& \qquad +\frac{1}{2}g_{\mu\nu}(V-e^{2a \phi}F^2-e^{-2a \phi}H^2). 
\end{align}
By taking the same solution for flux eq.~\eqref{sol_Av} and the same ansatz for the static metric eqs.~\eqref{metric_diagonal}, \eqref{advanced_null}, we find that the dilaton $\phi$ and the metric functions $\chi$ and $f$ satisfy  
\begin{align}
\label{scalar_field_Eq}
& f\phi''+\left(f'-\frac{f}{2}\chi'-\frac{2f}{z} \right)\phi' = \frac{1}{2} z^2 \partial_\phi \left( V_{\rm eff} +\frac{V}{2 z^4} \right),   \\
\label{chi_field_Eq}
& \chi'  = 2z\phi'^2,  \\
\label{f_field_Eq}
& f'   =z^3 \left( V_{\rm eff} +\frac{V}{2 z^4} \right) +\frac{f}{z}(3+z^2\phi'^2) \,.
\end{align}
Again, the effective potential is given by eq.~\eqref{defeffectivepotential}. 
From the comparison between these and eqs.~\eqref{scalar_field_Eq_effectivep}-\eqref{f_Eq}, we can see that the net effect of 
the addition of the relevant operator is to replace the effective potential into the new combination of the potential
\be
\label{netpotential}
 V_{\rm eff} +\frac{2 \Lambda}{2 z^4}  \to V_{\rm eff} +\frac{V(\phi)}{2 z^4}.  
\ee 

In this paper, we consider the potential  
\begin{align}
\label{barepotential2}
&V(\phi ) = - 6 \cosh m \phi  \,, \quad \Lambda = -3 \,, \\
&V(\phi) - 2 \Lambda = - 3 m^2 \phi^2 + {\cal{O}}(m \phi)^4  \,.
\end{align}
Note that our convention is such that for $m^2 > 0$, the dilaton is tachyonic, and 
with this mass term, the asymptotic behavior of the dilaton at $z\to 0$ behaves  
\be 
\label{asymp_scalar}
\phi\simeq \alpha z^{\Delta_-}+\beta z^{\Delta_+}, \qquad \Delta_\pm= \frac{1}{2} \left( 3 \pm \sqrt{9-6m^2} \right) \,. 
\ee
This dilaton becomes tachyonic in large $m^2$, and we need to choose its value satisfying the Breitenlohner-Freedman stability bound \cite{BF}
\be
\label{BFbound}
m^2 \le \frac{3}{2}.  
\ee

%%%%%%%%%%%%%%%%%%%%%
\subsection{Generalized attractor conditions}
%%%%%%%%%%%%%%%%%%%%%

To understand how the attractor solution reviewed in \S 2 flows to a new extremal solution by the relevant deformation, we first study the behavior of 
our extremal black hole solution near the horizon. 

Since the net effect of the bare potential $V(\phi)$ appears in the combination of eq.~\eqref{netpotential}, 
we notice immediately that at the extremal horizon after the addition of $V(\phi)$ must satisfy 
\be
\label{cond_Veff1}
\left( V_{\rm eff} +\frac{V(\phi)}{2 z_h^4}  \right)|_{\phi=\phi_0} = 0  \quad \mbox{(at the extremal horizon $z=z_h$)}.  
\ee  
This is seen from eq.~\eqref{f_field_Eq} because at the extremal horizon, both $f$ and $f'$ vanish. 
We also need 
\be
\label{cond_Veff2}
\partial_\phi \left( V_{\rm eff} +\frac{V}{2 z_h^4} \right) |_{\phi=\phi_0} = 0  \quad \mbox{(at the extremal horizon $z=z_h$)} 
\ee
for the regular extremal horizon to exist, since otherwise, the right hand side of eq.~\eqref{scalar_field_Eq} becomes nonzero 
although $f$ and $f'$ vanish and this leads to divergence of $\phi'$. 
On top of that, for the stability of the extremal horizon, we need 
\be
\label{cond_Veff3}
\tilde{M}_0^2 := \frac{1}{2} \partial_\phi^2 \left( V_{\rm eff} +\frac{V}{2 z_h^4} \right) |_{\phi=\phi_0} > 0  \quad \mbox{(at the extremal horizon $z=z_h$)},  
\ee
otherwise, the scalar fluctuation grows at the horizon. 
These conditions restrict the parameter space. In the large charge limit, the condition \eqref{cond_Veff1} gives 
\be
\label{zhqrela}
z_h \propto Q^{-\frac{1}{2} } \,,
\ee
if $Q_A $ and $Q_B $ are the same order ${\cal{O}}(Q)$. 
Then, the condition \eqref{cond_Veff2} and the condition \eqref{cond_Veff3} give 
\be
\label{amrelationship}
a^2 \gtrsim m^2 \,. 
\ee  
This can be understood as follows; 
$V_{\rm eff}$ yields {positive} mass-squared of the order of $ {\cal{O}}(a^2 Q^2)$, on the other hand, $V(\phi)$ yields {negative} mass-squared of the order of $ {\cal{O}}(m^2/z_h^4)$. Using eq.~\eqref{zhqrela}, one obtains eq.~\eqref{amrelationship} for the condition \eqref{cond_Veff3} to be satisfied.

To proceed further, as already indicated from the above analysis, it is convenient to 
introduce the following function:  
\be 
\label{defG}
G(\phi,\,z):= V_{\rm eff}(\phi)+\frac{V(\phi)}{2 z^4}.  
\ee
Then, the solutions near the extremal horizon of eqs.~\eqref{scalar_field_Eq}, \eqref{chi_field_Eq}, and \eqref{f_field_Eq} are 
determined by the function $G$ and at the horizon, one need 
\begin{align}
\label{cond_G1}
G(\phi,\,z) |_{\phi=\phi_0 \,, z=z_h} =  \partial_\phi G(\phi,\,z) |_{\phi=\phi_0 \,, z=z_h} &= 0 \,, \\ 
\tilde{M}_0^2 := \frac{1}{2} \partial_\phi^2 G(\phi,\,z) |_{\phi=\phi_0 \,, z=z_h} &> 0 \,.
\label{cond_G2}
\end{align}
These conditions can be re-written as follows. 
From eq.~\eqref{cond_G1}, the location of the extremal horizon, $z_h$, 
can be written in terms of the effective potential $V_{\rm eff}$ and bare potential $V$ as  
\be
\label{generalizedhorizon}
 z_h^4 = - \frac{V  |_{\phi=\phi_0} }{2 V_{\rm eff}  |_{\phi=\phi_0} }\ \quad (\mbox{at the extremal horizon}) \,.
\ee
The attractor value of $\phi$, {\it i.e.,} the value of $\phi$ at the horizon $\phi = \phi_0$ can be determined as 
\begin{align}
\label{generalizedattractor}
  \partial_\phi G(\phi,\,z) |_{\phi   =\phi_0 \,, z=z_h} &= 0  \nonumber \\
\Leftrightarrow \quad \partial_\phi  V_{\rm eff} (\phi) |_{\phi=\phi_0} &=  \frac{V_{\rm eff}  |_{\phi=\phi_0} }{V |_{\phi=\phi_0} }  \, \partial_\phi  V (\phi)  |_{\phi=\phi_0}  \,,  \\
&   \mbox{(at the extremal horizon $z=z_h$)}. \nonumber 
\end{align}
Equivalently, 
\begin{align}
\label{generalizedattractor2}
 & \partial_\phi \left( \log {V_{\rm eff} (\phi) }\right)  |_{\phi=\phi_0}= \partial_\phi \left( \log {  V (\phi) } \right) |_{\phi=\phi_0} \,,  \\
& \qquad \qquad  \qquad \mbox{(at the extremal horizon $z=z_h$)}. \nonumber 
\end{align}
The eq.~\eqref{generalizedattractor}, or equivalently eq.~\eqref{generalizedattractor2}, is the condition for the attractor value $\phi_0$ in the presence of the bare potential $V(\phi)$. 

Similarly using eq.~\eqref{generalizedhorizon}, the condition eq.~\eqref{cond_G2} can also be written in terms of potential only as 
\begin{align}
\label{generalizedattractor3}
\tilde{M}_0^2 &=  \frac{1}{2}  \partial_\phi^2 V_{\rm eff} |_{\phi=\phi_0} - \frac{1}{2} \frac{V_{\rm eff}  |_{\phi=\phi_0}}{ V |_{\phi=\phi_0}}  \partial_\phi^2 {V}  |_{\phi=\phi_0} > 0  \,. \\
\quad& \mbox{(at the extremal horizon $z=z_h$)} \nonumber
\end{align}

We summarize the net effect of the relevant operator at the extremal horizon, which is induced by the bare potential eq.~\eqref{barepotential2} with eq.~\eqref{barepotential1} as follows: 
\begin{enumerate}
\item The attractor value\footnote{Even though $\phi$ is no longer a moduli field in the presence of a bare potential, we still call $\phi_0$ as {\it attractor value} since that value is fixed by the charges of the black hole and the parameter of the theory only, and cannot take continuous value.}  given by eq.~\eqref{generalizedattractor}, (or equivalently eq.~\eqref{generalizedattractor2}), is the generalization of the attractor mechanism analyzed in \cite{attractor}, and in the absence of the bare potential $V(\phi)$, one can set $V(\phi) = 2 \Lambda$, and then, they reduces to eq.~\eqref{attractorpoint}. 
\item Similarly eq.~\eqref{generalizedattractor3} is the generalization of eq.~\eqref{attractorcondition} in the presence of the bare potential $V(\phi)$. 
\item The extremal horizon is set by eq.~\eqref{generalizedhorizon}. Without bare potential, it is given by eq.~\eqref{horizoncondition}. 
\end{enumerate}
As in the case of non-supersymmetric attractors \cite{attractor}, the horizon value of the dilaton is set by the charges of the black hole and the parameter of the theory only and cannot be modified continuously. 

%%%%%%%%%%%%%%%%%%%%%%%%%%%%%%%
\subsection{Near horizon analysis}
%%%%%%%%%%%%%%%%%%%%%%%%%%%%%%%

Given the attractor value by eq.~\eqref{generalizedattractor}, with the extremal horizon determined as eq.~\eqref{generalizedhorizon}, it is straightforward to analyze the behaviour of various fields near the horizon. 
To analyze eq.~\eqref{f_field_Eq} near the horizon, it is convenient to double-expand $z^3 G(\phi,\,z)$ as a series of  
\begin{align}
\delta\phi := \phi-\phi_0 \,, \qquad 
\delta z :=- z+z_h \,,
\end{align}
as follows,  
\begin{align}
\label{G_expansion}
& z^3 G(\phi,\,z)
= - \eta \delta z - \zeta\delta z\delta\phi+\frac{1}{2}\xi (\delta\phi)^2
+\frac{1}{2} \kappa (\delta z)^2 \nonumber \\
&\qquad  \qquad  \qquad  +(\mbox{higher orders of}\,\, \delta\phi, \,\delta z), 
\end{align}
where for the expansion~\eqref{G_expansion}, we have defined coefficients as 
\begin{align}
 \eta &:=\frac{\partial (z^3 G)}{\partial z}|_{\phi=\phi_0,\,z=z_h}, \quad \,\, \zeta:=\frac{\partial^2  (z^3 G)}{\partial\phi\partial z}|_{\phi=\phi_0,\,z=z_h},\\
\xi &:=\frac{\partial^2  (z^3 G)}{\partial\phi^2}|_{\phi=\phi_0,\,z=z_h}, \quad \kappa :=\frac{\partial^2 (z^3 G)}{\partial z^2}|_{\phi=\phi_0,\,z=z_h},
\end{align}
and we used the conditions~\eqref{cond_G1}.  

Next, from the regularity at the horizon, one can conclude that 
\be 
\label{validity_expansion}
|\delta z| > |\delta\phi|^2
\ee
near the horizon. This can be seen as follows; near the horizon eq.~\eqref{chi_field_Eq} yields, 
\be
 \partial_z \chi = 2 z_h  \left( {\partial_z \delta \phi} \right)^2. 
\ee
By imposing that $\chi$ is finite at the horizon for the minimal condition to have a regular extremal horizon, 
one obtains 
\be
\label{validity_expansion2}
\delta \phi \propto (\delta z)^\lambda \,, \quad \lambda > \frac{1}{2} \,,
\ee
and this yields eq.~\eqref{validity_expansion}. Then the dominant contribution from the right hand side of eq.~\eqref{G_expansion} is the term proportional to $\delta z$. By substituting this into  eq.~\eqref{f_field_Eq}, one obtains  
\begin{align}
 f'= - \eta \delta z + {\cal{O}}(\mbox{higher order terms in}\,\, \delta z,\,\delta\phi)+\frac{f}{z}(3+z^2\phi'^2). 
\end{align}
and its linearized solution near the horizon
\be
 f\simeq \frac{\eta}{2}(z-z_h)^2 + {\cal{O}}(\mbox{higher order terms in}\,\, \delta z,\,\delta\phi) \,.
\ee
With these, the dilaton eq.~\eqref{scalar_field_Eq}, becomes 
\begin{align} 
\label{linear_Eqs}
& \frac{\eta}{2}(z-z_h)^2  \delta \phi''+ \eta (z-z_h)  \delta \phi' -\frac{\xi}{2 z_h} \delta \phi 
=  \frac{\zeta}{2z_h}  (z-z_h)  \nonumber \\
&\qquad \qquad \qquad \qquad   + {\cal{O}}(\mbox{higher order terms}). 
\end{align}
This is a simple linear differential equation. Therefore the behavior of the dilaton near the horizon is obtained by summing both homogeneous solution $\phi_{H}$ and inhomogeneous solution $\phi_{IH}$ as 
\begin{align} 
&\phi \simeq  \phi_0 + \phi_{IH}+ C \phi_{H}  + {\cal{O}}(\mbox{higher order terms}) \,, 
\label{neigh_sol}
\\ 
& \phi_{H}:=(z_h-z)^\lambda,  \quad \lambda=\frac{-1+\sqrt{1+\frac{4\xi}{z_h\eta}}}{2} \,, 
\label{neigh_sol_H}
\\
& \phi_{IH} := \frac{\zeta}{2\eta z_h-\xi}(z-z_h) \,,
\label{neigh_sol_IH}
\end{align}
where $C$ is an arbitrary constant and the regularity at the horizon requires to choose positive root for $\lambda$. 
When $C\neq 0$, the inequality~\eqref{validity_expansion2} is equivalent to 
the condition 
\be 
\label{lambda_lower_bound}
\lambda>\frac{1}{2} \quad \Longleftrightarrow \quad \frac{4\xi}{z_h\eta}=4a^2-m^2>3. 
\ee   

Note that $\phi_{IH}$ corresponds to a smooth solution, while $\phi_{H}$ corresponds to 
 a class of $C^k$, with $k = \left \lfloor{ \lambda }\right \rfloor$, here $\left \lfloor{x}\right \rfloor$ is a floor function. 
 This is simply because $\lambda$ is not integer in general.   
In particular, when 
\be
\label{Upper_lambda}
\frac{1}{2}<\lambda<1,    
\ee
a parallelly propagated (p.~p.) curvature singularity\footnote{We can say that spacetime is singular when 
some curvature component for a parallelly propagated frame along a causal geodesic diverges indefinitely. This is 
called p.~p.~curvature singularity.} appears on the horizon, although all scalar curvature polynomials such as 
$R^{\mu\nu\alpha\beta}R_{\mu\nu\alpha\beta}$ are finite there. In the null coordinate system given by eq.~\eqref{advanced_null}, 
one can consider the following affine-parametrized radial null geodesic with tangent vector 
\be
l=\partial_\lambda=\frac{dz}{d\lambda}\partial_z=z^2e^{\frac{\chi}{2}}\partial_z.  
\ee
Then, we find that Ricci curvature component $R_{\mu\nu}l^\mu l^\nu$ in the parallelly propagated frame 
diverges under the condition eq.~\eqref{Upper_lambda} in the limit $z\to z_h$ as
\be 
R_{\mu\nu}l^\mu l^\nu=z^4e^{\chi}R_{zz}=2z^4\phi'^2e^\chi\sim C^2{\phi'}_{H}^2\sim C^2(z_h-z)^{2\lambda-2}\to \infty ,  
\ee
where we used the fact that $\chi$ is finite for eq.~\eqref{Upper_lambda} via eq.~\eqref{chi_field_Eq}. 
Note that, however, 
this is a {\it mild} singularity in the sense that the integral of $R_{\mu\nu}l^\mu l^\nu$ is finite, {\it i.e.}, 
$\int R_{\mu\nu}l^\mu l^\nu d\lambda<\infty$ in the parameter range eq.~\eqref{Upper_lambda}. 
This means that the expansion $\theta:=\nabla^\mu l_\mu$ of the null geodesic congruence along the radial null geodesic, 
which is one of the fundamental quantities characterizing singularity, is finite on the extremal horizon 
via the Raychaudhuri equation
\be 
\frac{d\theta}{d\lambda}=-\frac{1}{2}\theta^2-R_{\mu\nu}l^\mu l^\nu. 
\ee 
Therefore, the p.~p.~singularity is quite different from the strong curvature singularity appearing in the 
dilatonic extremal black holes with no attractor mechanism, in which the dilaton diverges on the extremal 
horizon~\cite{Gibbons_Maeda1988, GHS1991}. 
The p.~p.~curvature singularity was also found in extremal inhomogeneous Reissner-Nordstrom 
AdS solution~\cite{MaedaOkamuraKoga}.  

%%%%%%%%%%%%%%%%%%%%%
\subsection{Interpolating near horizon to asymptotic AdS}
%%%%%%%%%%%%%%%%%%%%%
Given the effective potential eq.~\eqref{defeffectivepotential} and the bare potential eq.~\eqref{barepotential2},
the condition eq.~\eqref{generalizedhorizon} becomes
\be
\label{cond_zh}
z_h^4 =\frac{3\cosh m \phi_0}{Q_A^2e^{-2a \phi_0}+Q_B^2e^{2a \phi_0}} \,, 
\ee
where $\phi_0$ is the value at the extremal horizon. 
The condition eq.~\eqref{generalizedattractor} becomes 
\begin{align}
\label{cond_V_e}
Q_B^2 &=e^{-4a \phi_0}\frac{m\sinh m\phi_0+2a \cosh m\phi_0}
{2a \cosh m\phi_0-m\sinh m\phi_0} \, Q_A^2 \,, 
\end{align}
and the condition \eqref{generalizedattractor3}
becomes 
\be
2 a > m \,.
\ee
Note that $\phi_0=0$ when $Q_A=Q_B=Q$, and 
$Q_B/Q_A \to 0$ as $\phi_0\to\infty$~(for  $\phi_0\to-\infty$, $Q_A/Q_B \to 0$). 
So, $\phi_0$ represents deviation from $|Q_B-Q_A|=0$. 

As described in the previous subsection, a {p.~p.~singularity} appears on the extremal horizon when $\lambda$ 
satisfies \eqref{Upper_lambda}. So, in principle, it would be possible to describe such a {p.~p.~singularity} by the dual field theory 
via the AdS/CFT dictionary, since the singularity appears on the boundary of the causal wedge of the whole boundary spacetime. 
In the non-extremal black hole case, the geometric property of singularity inside the event horizon was vigorously 
investigated in~\cite{FHKS2020, HHKS2020, HHKS2021}, but whether such a singularity inside the black hole can be 
described by the dual field theory is still not clear. Therefore, partially motivated by the above expectation, in this paper, we shall pay 
attention to the parameters in the range of $\lambda$ in \eqref{Upper_lambda}, 
\be 
\label{amqa}
a=\frac{6}{5}, \qquad m=\frac{2}{\sqrt{3}}, \qquad Q_A=1 \quad \Longrightarrow \quad \lambda\simeq  0.66.  
\ee
Then, both $\Delta_+$ and $\Delta_-$  become integers and both of the mode functions $z^{\Delta_+}$ and $z^{\Delta_-}$ are normalizable. 
In this case, one can consider a generalized boundary condition at the 
AdS boundary,  
\be
\label{generalized_bc}
\beta(\alpha)=\beta_{\rm b.c.}(\alpha), 
\ee
where $\beta_{\rm b.c.}(\alpha)$ reflects our choice for the boundary condition. 
In this paper, we will choose our boundary condition as  
\begin{align}
\label{ourbccondition}
\beta_{\rm b.c.}(\alpha)= k\alpha (\alpha-\alpha_0), 
\end{align}
where $k$ and $\alpha_0$ are some positive constants\footnote{When $\alpha$ is small, this corresponds to 
the Robin condition
\begin{align}
\p_z f\propto f, 
\end{align}
where $f=\phi/z$~\cite{Bizon2020}. When $\alpha$ is large enough, the boundary condition reduces to 
$\beta\simeq k\alpha^2$ that preserves all the asymptotic AdS symmetries~\cite{HertogMaeda2004}.}.

Under the boundary condition~\eqref{ourbccondition}, one can numerically find the solutions of 
eqs.~\eqref{scalar_field_Eq}, \eqref{chi_field_Eq}, and \eqref{f_field_Eq}, which interpolate near horizon to 
asymptotic AdS boundary. 

We can also tune parameters in such a way that  
\be
\phi_0 = 1  \,,
\ee
then correspondingly, the other parameters are set as  
\be
z_h \simeq 2.044 \,, \qquad Q_B^2 \simeq  0.01894 \,
\ee
from eq.~\eqref{cond_zh} and \eqref{cond_V_e}. 

The asymptotic behaviors of $\phi$, $\chi$, and $f$ become 
\begin{align} 
\label{asymp_sol}
& \phi\simeq \alpha z+\beta z^2, \\
& \chi\simeq\alpha^2z^2+\frac{8}{3}\alpha\beta z^3+2\beta^2 z^4+\cdots, \nonumber \\ 
& f\simeq 1+\alpha^2 z^2-Mz^3+(\alpha^4+2\beta^2)z^4+\cdots. \nonumber 
\end{align}
As seen in the solution \eqref{neigh_sol} near the horizon, there are two choices on the boundary condition at the 
extremal horizon, (i) $C=0$ smooth solution and (ii) $C \neq 0$ {p.~p.~singular non-smooth solution}. 

\begin{figure}[htbp]
  \begin{center}
   \includegraphics[width=100mm]{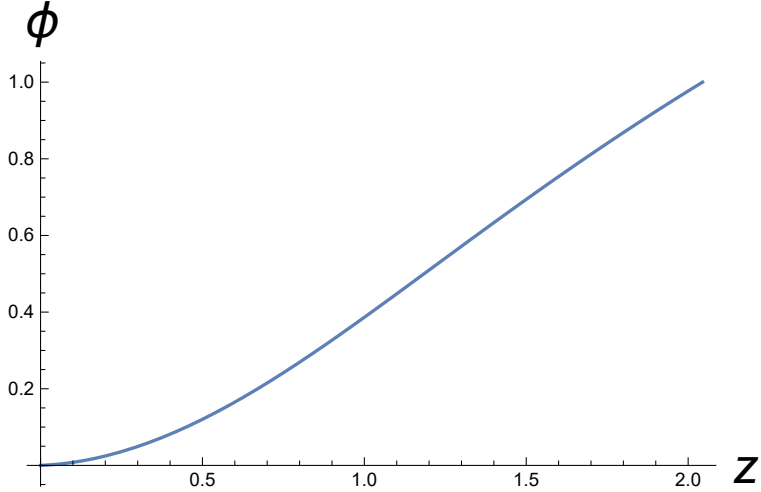}
  \end{center}
  \caption{The dilaton profile $\phi(z)$ interpolating the horizon  $z = z_h \simeq 2.044$ to the AdS boundary $z=0$ for 
 $a=6/5$, $m=2/\sqrt{3}$, $Q_A=1$, $\phi_0=1$, and $C=0$.}
  \label{phi_phi=1_Extremal_BH}
 \end{figure}

\begin{figure}[htbp]
  \begin{center}
   \includegraphics[width=100mm]{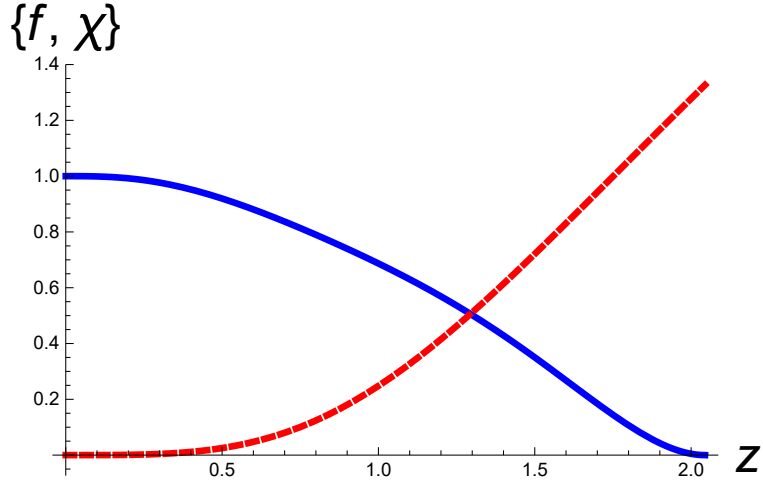}
  \end{center}
  \caption{The warped factor $f$~(solid blue curve) and $\chi$~(dashed red curve) in the metric interpolating 
  the horizon $z = z_h \simeq 2.044$ to the AdS boundary $z=0$ for 
  $a=6/5$, $m=2/\sqrt{3}$, $Q_A=1$, $\phi_0=1$, and $C=0$.}
  \label{f_phi=1_Extremal_BH} 
\end{figure}

Figure \ref{phi_phi=1_Extremal_BH} and \ref{f_phi=1_Extremal_BH}  show the numerical solution for the case (i), in which 
$\phi_0=1$ and $Q_A=1$. The solution is everywhere smooth. 
%%%%%%%%%%%%%%%%%%%%%%%%%%%%%%%%%%%%%%%
%%%%%%%%%%%%%%%%%%%%%%%%%%%%%%%%%%%%%%%

\begin{figure}[htbp]
 \begin{center}
 \includegraphics[width=100mm]{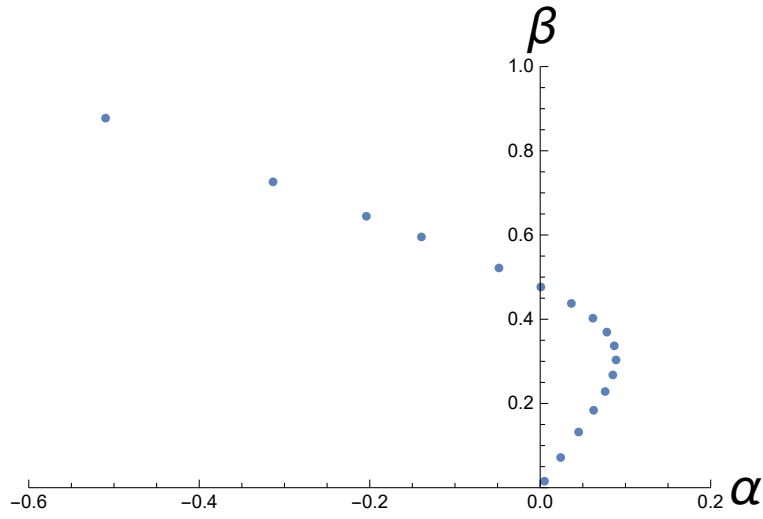}
 \end{center}
 \caption{$\alpha$-$\beta$ plot for $a=6/5$, $m=2/\sqrt{3}$, $Q_A=1$, 
 $C=0$ by varying $Q_B$. We vary $Q_B$  in such a way that through eq.~\eqref{cond_V_e}, the resultant  $\phi_0$ changes from $0$ to $1.6$.}
 \label{c1=1_alpha_beta_relation}
\end{figure}

One can also consider one-parameter family of solutions by varying $Q_B$.  
Given $a$, $m$, $Q_A$ fixed by eq.~\eqref{amqa}, varying $Q_B$ corresponds to varying $\phi_0$ through eq.~\eqref{cond_V_e}. 
We will fine-tune and vary $Q_B$ in such a way that through eq.~\eqref{cond_V_e}, the resultant  $\phi_0$ changes from $0$ to $1.6$.  Eq.~\eqref{cond_zh} sets the horizon $z_h$ accordingly.  We keep $C=0$  in eq.~\eqref{neigh_sol}, so that homogeneous non-smooth solutions do not contribute. 
Then, by extrapolating the near horizon solution to asymptotic AdS boundary, one can read off the dilaton profile at the boundary, which is parametrized by $\alpha$ and $\beta$ as eq.~\eqref{asymp_sol}. How $\alpha$ and $\beta$ are related in this one-parameter family of solution is shown in Fig.~\ref{c1=1_alpha_beta_relation}. 

%%%%%%%%%%%%%%%%%%%%%%%%%%%%%%%%%%%%%%%
%%%%%%%%%%%%%%%%%%%%%%%%%%%%%%%%%%%%%%%%

%%%%%%%%%%%%%%%%%%%%%%%%%%%%%%%%%%%%%%%%%
\begin{figure}[htbp]
 \begin{center}
   \includegraphics[width=100mm]{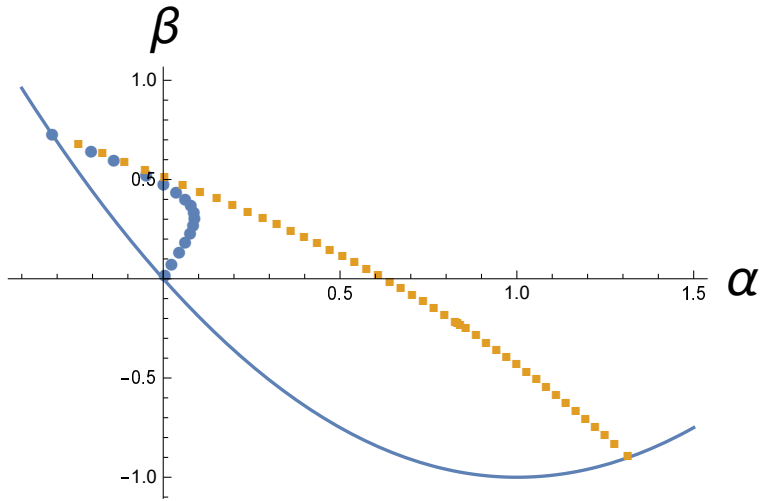}
 \end{center}
 \caption{The orange squares represent p.~p.~singular  
 solutions for $a=6/5$, $m=2/\sqrt{3}$, $Q_A=1$, $Q_B = 0.046$, $\phi_0=1.49$, and by varying $C$ from 0 (left) to 0.22 (right). We call this orange square curve as $\beta_s(\alpha)$. 
 The solid curve, on the other hand, represents $\beta=k \alpha(\alpha- \alpha_0)$, where $k=1$, and $\alpha_0 =2$. The blue circle represents the solutions for various $Q_B$, with $a=6/5$, $m=2/\sqrt{3}$, $Q_A=1$, 
 $C=0$ as Fig.~\ref{c1=1_alpha_beta_relation}.}
 \label{c1=1_phi0_149_NonAnalytic}
\end{figure}
%%%%%%%%%%%%%%%%%%%%%%%%%%%%%%%%%%%%%%%%

For a fixed $Q_A$, $Q_B$, $a$, $m$, and accordingly $\phi_0$ and $z_h$, one can also consider (ii) $C \neq 0$ p.~p.~singular non-smooth solutions by varying $C$ from $C=0$ smooth solution. See Fig.~\ref{c1=1_phi0_149_NonAnalytic}, where we  
show one parameter family of p.~p.~singular non-smooth solutions by varying $C$ while keeping $a=6/5$, $m=2/\sqrt{3}$, $Q_A=1$, $Q_B = 0.046$, $\phi_0=1.49$ fixed,  in the orange squares.  We call this orange-square curve  as 
\be
\beta(\alpha) = \beta_s(\alpha) \,. 
\ee

So far we have not considered the boundary condition but now we consider it. From eq.~\eqref{ourbccondition}, when
\be 
\label{intersection_beta}
\beta_s(\alpha) = \beta_{\rm b.c.} (\alpha)= k\alpha (\alpha-\alpha_0),   
\ee  
is satisfied, we can obtain the solutions obeying the boundary condition. 
For example, $k=1$ and $\alpha_0=2$ case is plotted in Fig.~\ref{c1=1_phi0_149_NonAnalytic}.
There are two solutions satisfying this boundary condition. The left intersection 
point represents a smooth solution $C=0$, while the right intersection point represents a p.~p.~singular non-smooth solution $C=0.22$ for the same charges $Q_A$ and $Q_B$. Given these two solutions with the same charges, it is natural to ask which one is more stable.

%%%%%%%%%%%%%%%%%%%%%%%%
\section{Stability of the solutions} 
\label{sec:4} 
%%%%%%%%%%%%%%%%%%%%%%%%
%%%%%%%%%%%%%%%%%%%%%%%%%%%%%%%%%%%%%%%%%
\subsection{Bulk interpretation}
%%%%%%%%%%%%%%%%%%%%%%%%%%%%%%%%%%%%%%%%%

Finding extremal black hole solutions, interpolating from the horizon to the AdS boundary, we would like to study its thermodynamical stability. 
Since the black hole we find in the bulk is extremal black hole, their temperature vanishes.  
In this case, the free energy reduces to simply the total energy of the black hole, which can be derived from the Hamiltonian approach.  For later convenience, let us compactify both $x$ and $y$ direction 
with length $L$. 
Let $\xi^\mu$ be a timelike vector which asymptotically approaches time translation in asymptotic AdS. The variation 
of the gravitational surface term is given by \cite{RT74}
\begin{align}
\label{energy_G} 
& \delta Q_G[\xi]=\frac{1}{2}\int dxdy \delta^z_i\,\overline{G}^{ijkl}(\xi_\perp \bar{D}_j\delta h_{kl}-\delta h_{kl}\bar{D}_j \xi^\perp), 
\nonumber \\
& G^{ijkl}=\frac{1}{2}\sqrt{g}(g^{ik}g^{jl}+g^{il}g^{jk}-2g^{ij}g^{kl}), \qquad h_{ij}=g_{ij}-\bar{g}_{ij}. 
\end{align} 
Here, $\bar{g}_{ij}~(i,\,j=x,y,z)$ is the spatial metric of pure AdS and $\bar{A}$ represents 
the quantity evaluated by $\bar{g}_{ij}$, and $\xi_\perp=\xi^\mu n_\mu$, where $n^\mu$ is the unit normal to the Cauchy 
surface $\Sigma$. The variation of the surface term by scalar field is also given by   
\be 
\label{energy_scalar}
\delta Q_\phi=-\int \xi^\perp \delta\phi D_i\phi \,dS^i. 
\ee
In general, both eq. \eqref{energy_G} and \eqref{energy_scalar} diverge, due to the slower fall-off of the scalar 
field with $\beta=\beta(\alpha)$. By using the asymptotic expansion \eqref{asymp_sol}, they are evaluated as
\be
\label{energy_gravity_express} 
\delta Q_G=\frac{L^2}{2}\left(\delta M-\frac{\delta(\alpha^2)}{z}\right), 
\ee
\be 
\label{energy_scalar_express}
\delta Q_\phi=L^2\left[\frac{1}{2z}\delta(\alpha^2)+\delta(\alpha\beta)+\beta\delta\alpha \right]. 
\ee   
The divergent term in the gravitational energy~\eqref{energy_gravity_express} is exactly cancelled by the divergent term 
in the energy~\eqref{energy_scalar_express} contributed by the scalar field as   
\begin{align}
\delta Q=\delta Q_G+\delta Q_\phi=L^2\left(\frac{\delta M}{2} +\delta(\alpha\beta)+\beta\delta\alpha  \right). 
\end{align}
Therefore, under the generalized boundary condition~\eqref{generalized_bc}, by integrating $\delta Q$ under the relation $\beta=\beta_{\rm b.c.}(\alpha)$ 
we obtain the following total energy, 
\begin{align} 
\label{total_energy}
E=L^2\left(\frac{M}{2}+\alpha\beta+W(\alpha)  \right), \qquad  
W(\alpha):=\int^\alpha_0 \beta_{\rm b.c.}(\tilde{\alpha})d\tilde{\alpha},    
\end{align}
where $M$ is the mass term appeared in the asymptotic expansion of $f$. 
As shown below, using eq.~\eqref{total_energy}, we discuss the stability of the bulk solutions.   

The total energy $E$ in~\eqref{total_energy} is $E\simeq 0.68L^2$ for (i) the smooth 
solution, while $E\simeq -0.25L^2$ for (ii) the p.~p.~singular non-smooth solution in the choice $k=1$ and $\alpha_0=2$ in 
eq.~\eqref{ourbccondition}. For another choice, $k=2$ and $\alpha_0=2$,  
we can also find two characteristic solutions (i) and (ii) for $\phi_0\simeq 1.33$ and the 
total energy for the p.~p.~singular solution is lower than the one of smooth solution, {\it i.e.}, 
$E\simeq 0.65L^2$ for (i), and $E\simeq -1.31L^2$ for (ii).  This characteristic 
behavior seems to be independent of the choice of the parameters, $k$ and $\alpha_0$. 
Therefore the energy for the p.~p.~singular non-smooth solution is always lower than the one of the smooth solution under the 
boundary condition~(\ref{ourbccondition}). 
This implies that the non-smooth solution with nonzero $C$ is more stable than the one for smooth solution with $C=0$.

%%%%%%%%%%%%%%%%%%%%%%%%%%%%%%%%%%%%%%%%%
\subsection{Boundary interpretation}
%%%%%%%%%%%%%%%%%%%%%%%%%%%%%%%%%%%%%%%%%

Now, we can investigate our bulk solutions from the viewpoint of the dual field theory. 
Let $S_0$ denote the Lagrangian in the boundary field theory. As shown in \cite{Witten01}, the generalized boundary 
condition~\eqref{generalized_bc} corresponds to the deformation 
\be 
\label{multi-trace}
S=S_0+\int W({\cal O}), 
\ee
where $W({\cal O})$ is the function of the dimension one scalar operator ${\cal O}$ 
dual to the scalar field $\phi$, as defined in \eqref{total_energy}. 
More concretely in our choice of the boundary condition eq.~\eqref{ourbccondition}, 
\be
W({\cal O})= \frac{k}{3} {\cal O}^3 - \frac{k \alpha_0}{2}  {\cal O}^2 \,.
\ee
The stability can be described by a potential ${\cal V}$ \cite{HH05} as follows.  

Let us define a function $W_0(\alpha)$ and ${\cal V}(\alpha)$ as 
\be
\label{def:W_0} 
W_0(\alpha)=-\int^\alpha_0 \beta_s(\tilde{\alpha})d\tilde{\alpha}, \qquad 
{\cal V}(\alpha):=W(\alpha)+W_0(\alpha),  
\ee
where $\beta_s(\alpha)$ is the one-parameter bulk solution shown in Fig.\ref{c1=1_phi0_149_NonAnalytic} by orange square curve. 
Then, the solution for a boundary condition 
$\beta=\beta_{\rm b.c.}(\alpha)$ is given by the extremality condition of $\cal V$,  
\be 
0=\partial_\alpha{\cal V}=\partial_\alpha W+\partial_\alpha W_0=\beta_{\rm b.c.}-\beta_s,  
\ee 
agreeing with eq.~\eqref{intersection_beta}. As argued in  
\cite{HH05}\footnote{In \cite{HH05}, the gravitational soliton 
solutions were numerically constructed under the boundary condition $\beta=-k$. The scalar hairy 
black hole solutions were also constructed under another boundary condition, 
$\beta=-k\alpha^2+\epsilon \alpha^3$, where $\epsilon$ is introduced to produce a stable ground 
state \cite{HertogHorowitzJHEP2005}. In either case, two solutions were found for each boundary condition $\beta=\beta(\alpha)$ for a certain parameter range of $k$ and $\epsilon$.}, the solution is stable 
when the extrema is minimum, 
{\it i.e}., $\partial_\alpha^2{\cal V}>0$, while it is unstable when the extrema is maximum, {\it i.e.}, $\partial_\alpha^2{\cal V}<0$. 
As shown in Fig.~\ref{c1=1_phi0_149_NonAnalytic}, there are two solutions for the 
boundary conditions given by eq.~(\ref{ourbccondition}) for various $k$ and $\alpha_0$. For $k=1$ and $\alpha_0=1$ case, 
$\partial_\alpha^2{\cal V}<0$ for $\alpha\simeq -0.313$ smooth solution, and $\partial_\alpha^2{\cal V}>0$ for $\alpha\simeq  1.31$ 
non-smooth solution. 
Therefore this potential analysis suggests that the non-smooth solution is stable, while the smooth solution is unstable. 
This is indeed consistent with the fact that the energy of the former solution is less than that of the latter smooth solution found 
in the previous subsection.

%%%%%%%%%%%%%%%%%%%%%%%%
\section{Summary and discussions} 
\label{sec:5} 
%%%%%%%%%%%%%%%%%%%%%%%%

In this paper, we have addressed, in the holographic context, the question of how extremal black holes flow, when a relevant deformation is introduced to the bulk theory, to new extremal black holes. For this purpose, we have considered extremal black holes with a single dilaton field coupled to two Maxwell fields in four-dimensional asymptotically AdS spacetimes, which exhibit the non-supersymmetric attractor phenomenon.

In section~\ref{sec:2}, we review the attractor mechanism and we have perturbatively constructed an attractor solution, starting 
from the extremal Reissner-Nordstrom AdS black hole with planar horizon as our zero-th order solution. 
Then, in section~\ref{sec:3}, we have examined the flow of the attractor black holes by making the relevant deformation of the moduli potential to the bulk action. This has been done by adding the bare potential for the dilaton field.  
The bare potential induces the bulk to new extremal black hole solution. 
The net effects of the relevant deformation at the new extremal horizon can be summarized as eq.~\eqref{generalizedhorizon}, \eqref{generalizedattractor}, \eqref{generalizedattractor2}, \eqref{generalizedattractor3}, which reduces to eq.~\eqref{horizoncondition}, \eqref{attractorpoint}, \eqref{attractorcondition} respectively in the absence of the bare potential. 
Especially, as we have found that in the case of \cite{attractor}, the horizon values of the dilaton is set by the black hole charges and the parameters of the theory only as eq.\eqref{generalizedattractor}, or equivalently \eqref{generalizedattractor2}. Furthermore, we have numerically found the global solutions which interpolate between the near horizon and the AdS asymptotic regions. 

We have examined the asymptotic behavior of the dilaton field and the metric functions. 
The near horizon analysis has been done by expanding the term $z^3 G(\phi,z)$ in terms of both the dilation field $\phi$ and the bulk radial coordinate $z$ as eq.~\eqref{G_expansion}. 
It turned out that the near horizon dilaton solution consists of the two parts: the smooth solution $\phi_{IH}$ given by (\ref{neigh_sol_IH}) and the non-smooth one, $\phi_{H}$, by (\ref{neigh_sol_H}). 
The latter can give rise to a p.~p.~singularity at the horizon under the parameter region given in eq.~\eqref{Upper_lambda}. 
We would like to stress that the appearance of the smooth solution $\phi_{IH}$ is one of the new effects introduced by our relevant deformation, which causes the flow of the attractor black holes. 

As for the asymptotic region, the relevant deformation corresponds to changing boundary conditions at the AdS boundary to the generalized conditions~(\ref{generalized_bc}), admitting the two types of normalizable modes $z^{\Delta_\pm}$ near the AdS boundary. 
% For certain parameter values, we have numerically found the solutions that interpolate 
% between the near horizon region and the AdS asymptotic region. 
As just mentioned above, 
depending upon the boundary conditions at the horizon, we have obtained smooth global solutions and non-smooth ones.  

Having obtained the global solutions, we have studied, in section~\ref{sec:4}, their thermodynamic stability from the bulk theory viewpoint by examining their free energy. Since our solutions have zero-temperature, their free energy is equivalent to the total energy, which we can evaluate as the total Hamiltonian. We have found that the total energy for 
the non-smooth solution is always lower than that for the smooth solution, thus implying that the non-smooth solution is thermodynamically more stable than the smooth solution. Note that although we have not analyzed all possible cases, 
our stability result appears to be irrespective of the choice of the generalized boundary conditions~(\ref{generalized_bc}).

Since our relevant deformation corresponds, in the context of gauge/gravity duality, to adding a relevant operator to the boundary field theory, it is 
also natural to examine the stability of our solution from the boundary field theory perspective. We have found that the non-smooth solution is more 
stable than the smooth one, in agreement with the stability result from the bulk perspective.

It would be interesting to study dynamical stability of our attractor solutions, 
in particular those of smooth, thermodynamically unstable solutions. It has been known from linear analysis~\cite{AIWald04} that pure AdS spacetime can be unstable for certain mixed, linear boundary conditions, which may be viewed as a (linear version of the) relevant deformation. We may therefore expect that similar to the linear analysis, our smooth solutions can be dynamically stable or unstable depending on the choice of boundary conditions at the AdS boundary. We should however note that our boundary conditions (\ref{generalized_bc}) are non-linear, and hence it is non-trivial whether our thermodynamically unstable solutions can also become dynamically unstable.

In this paper, we have focused on the planar horizon black hole with a single dilaton field in the parameter range where a p.~p.~singularity exists, namely $1 > \gamma > 1/2$. 
It would be interesting to generalize our present analysis to $\gamma > 1 $ case as well, where there is no p.~p.~ singularity and study their stability and check if non-smooth attractor-like $C \neq 0$ solution is generically stable or not. 
It would also be interesting to generalize to the system including more moduli fields as well as the spherical horizon. Finally it is interesting to investigate if there is a way to see the signal of the p.~p.~singularity from the boundary dual, possibly through the entanglement entropy.  We hope to come back to these questions in the near future.

\bigskip
\goodbreak
\centerline{\bf Acknowledgments}
\noindent

This work was supported in part by JSPS KAKENHI Grant No. 18K03619 (N.I.), 15K05092 (A.I.), 20K03975 (K.M.) and also supported by MEXT KAKENHI Grant-in-Aid for Transformative Research Areas A Extreme Universe No.21H05184 (N.I.), and No.21H05186 (A.I. and K.M.). 

%\appendix
%%%%%%%%%%%%%%%%%%%%%%%%%%%%%%%%%%%%%%%%%%%%%%%%%%%%%%%%%%%%%%%%%%%%%%%%%%%%%%
%\section{The total finite energy}
%\label{Appendix:A}
%%%%%%%%%%%%%%%%%%%%%%%%%%%%%%%%%%%%%%%%%%%%%%%%%%%%%%%%%%%%%%%%%%%%%%%%%%%%%%

%%%%%%%%%%%%%%%%%%%%%%%%%%%%%%%%%%%%%%%%%%%%%%%%%%%%%%%%%%%%%%%%%%%%%%%%%%%%%%%%

\end{document}